\documentclass[journal=jacsat,manuscript=article]{achemso}
\setkeys{acs}{articletitle = true}

\usepackage[version=3]{mhchem} 

\usepackage{xcolor}

\author{Crist\'{o}bal P\'{e}rez}\email{cristobal.perez@ehu.es}\affiliation{Deutsches Elektronen-Synchrotron DESY, Notkestra{\ss}e 85, D-22607 Hamburg, Germany}\alsoaffiliation{Facultad de Ciencia y Tecnolog\'{i}a, Universidad del Pa\'{i}s Vasco (UPV-EHU), E-48940 Leioa, Spain}\alsoaffiliation{Ikerbasque, Basque Foundation for Science, E-48013 Bilbao, Spain}
\author{Amanda L. Steber}\affiliation{Deutsches Elektronen-Synchrotron DESY, Notkestra{\ss}e 85, D-22607 Hamburg, Germany}\alsoaffiliation{Christian-Albrechts-Universit{\"a}t zu Kiel, Institute of Physical Chemistry, Max-Eyth-Stra{\ss}e 1, D-24118 Kiel, Germany}\alsoaffiliation{The Hamburg Centre for Ultrafast Imaging at the University of Hamburg, D-22761 Hamburg, Germany}
\author{Anna Krin}\affiliation{Deutsches Elektronen-Synchrotron DESY, Notkestra{\ss}e 85, D-22607 Hamburg, Germany}\alsoaffiliation{Christian-Albrechts-Universit{\"a}t zu Kiel, Institute of Physical Chemistry, Max-Eyth-Stra{\ss}e 1, D-24118 Kiel, Germany}\alsoaffiliation{The Hamburg Centre for Ultrafast Imaging at the University of Hamburg, D-22761 Hamburg, Germany}
\author{Melanie Schnell}\email{melanie.schnell@desy.de}\affiliation{Deutsches Elektronen-Synchrotron DESY, Notkestra{\ss}e 85, D-22607 Hamburg, Germany}\alsoaffiliation{Christian-Albrechts-Universit{\"a}t zu Kiel, Institute of Physical Chemistry, Max-Eyth-Stra{\ss}e 1, D-24118 Kiel, Germany}\alsoaffiliation{The Hamburg Centre for Ultrafast Imaging at the University of Hamburg, D-22761 Hamburg, Germany}

\title{State-specific Enrichment of Chiral Conformers with Microwave Spectroscopy}
     
\abbreviations{}

\begin{document}




\begin{abstract}
An interesting class of molecules is that in which the molecules do not possess a stereogenic center but can become chiral due to their spatial arrangement. These molecules can be seen as chiral conformers, whose two non-superimposable forms can interconvert from one another by rotations about single bonds. Here, we show that an initially racemic mixture of chiral conformers, such as a sample of cyclohexylmethanol, C$_{7}$H$_{14}$O (CHM), can be enantiomerically enriched by performing the enantio-selective process of coherent population transfer between rotational levels. By first performing a population transfer cycle, followed by a three-wave mixing experiment, we show that an enantiomeric excess in a rotational level of choice can be achieved. This represents the first experimental demonstration of such an effect in a chiral pair of conformers, and it showcases the broad applicability of three-wave mixing not only for analytical applications but also to a wide scope of experiments of fundamental interest.           

\end{abstract}

\begin{figure}[h] 
	\centering
	\includegraphics[scale=1]{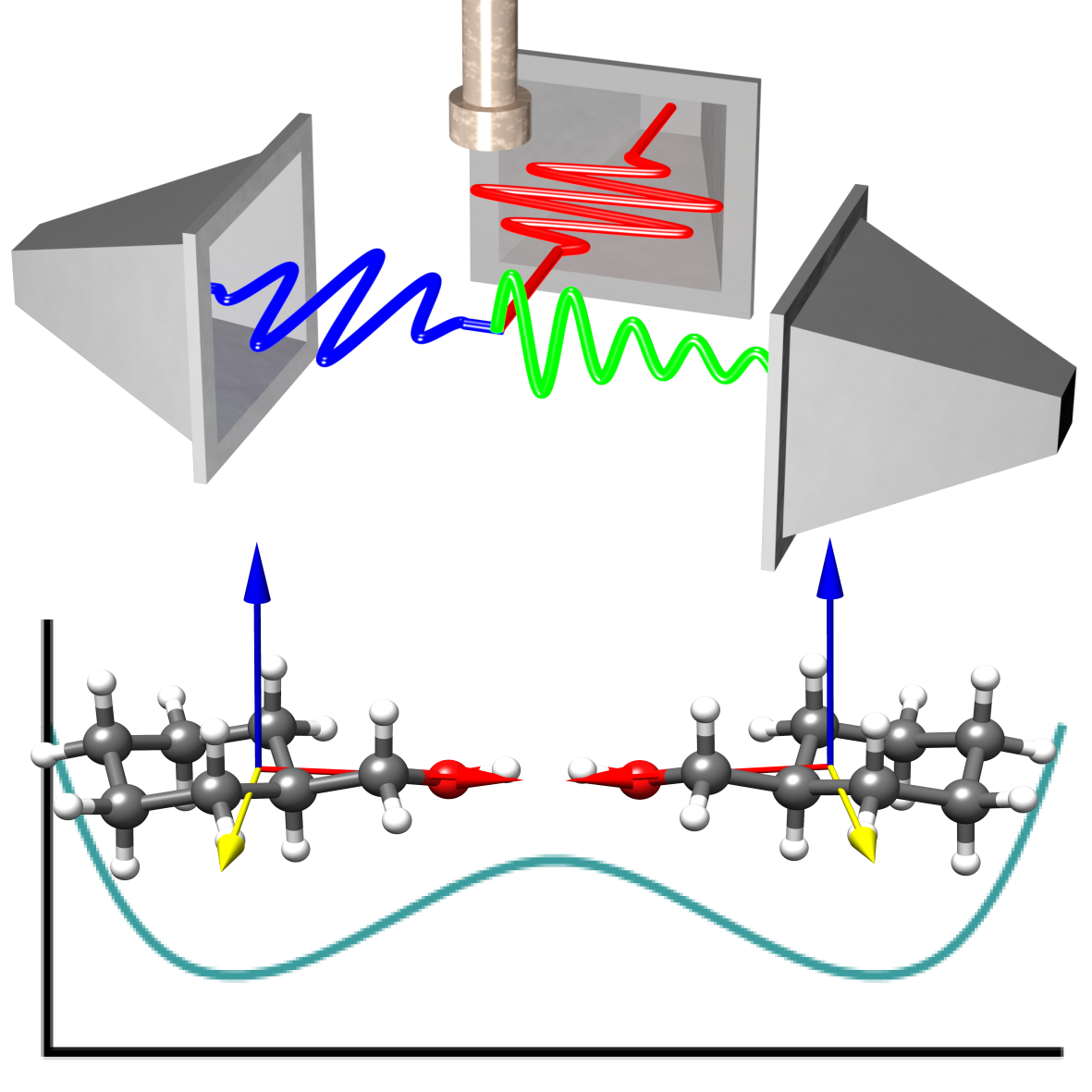} 
\end{figure}



Chirality is a topic of interest in many areas of research ranging from physics, chemistry, and biochemistry. In the simplest scenario, a chiral molecule has two enantiomers that share many physical and chemical properties; however, the enantiomers are non-superimposable mirror-images whose functionality is often dominated by the environment in which they are present. This is especially relevant when talking about enantiomer functionality in the body, as one enantiomer may be active while the other may not. Because of this, molecular handedness is a decisive factor in modern asymmetric synthesis, biotechnology and pharmaceutical production. The origin of this biological enantioselectivity, first suggested by Pasteur in 1854,\cite{pasteur_sur_1948} is still disputed, but it may eventually be traced to fundamental aspects of physics, like parity violation.\cite{quack_high-resolution_2008} Despite its undeniable importance, the detection and in particular the quantification of chirality remain challenging issues. A key step in understanding an enantiomer's functionality relies on being able to accurately determine its specific handedness (absolute configuration), enantiomeric excess (ee), conformation, and structural heterogeneity at a molecular level.

The chiroptical techniques available for the determination of the absolute configuration and ee of a chiral sample depend on the fact that the symmetry of enantiomers can be broken using circularly-polarized radiation. Circular Dichroism, Vibrational Circular Dichroism (VCD)\cite{NafieBOOK} and Raman Optical Activity (ROA)\cite{Barron2007} have emerged as the dominant methods to determine the absolute configuration of chiral molecules by comparison with quantum-chemical calculations. These techniques are, however, at times limited due to the intrinsic mechanism that governs their intensity (the interference between the electric transition-dipole moment and the weak magnetic transition-dipole moment). Among other recent developments, Photoelectron Circular Dichroism (PECD), which uses intense laser light and/or synchrotron sources,\cite{lux_circular_2012-1,garcia_vibrationally_2013,Janssen2014,Horsch2012} and Coulomb Explosion with Coincidence Imaging\cite{pitzer_direct_2013-1,herwig_absolute_2014} are emerging as important techniques for tackling the difficult problem of chiral quantification.  

In order to try to solve some of the remaining issues in analyzing chiral molecules, the microwave three-wave mixing (M3WM)\cite{patterson_enantiomer-specific_2013,patterson_sensitive_2013,shubert_identifying_2014,shubert_chiral_2016} approach was introduced a few years ago, and since then work has been carried out to further improve its usefulness. While enantiomers share most of their physical properties with one another, they differ in the triple product $\vec{\mu}_a\cdot(\vec{\mu}_b\times\vec{\mu}_c)$ of the electric-dipole moment components in the molecular principal axis system.\cite{hirota_triple_2012,grabow_fourier_2013} This feature can be exploited by M3WM to discriminate a molecule's chirality by exciting closed loops of rotational levels that involve all three types of dipole-allowed transitions via a double resonance scheme with controlled polarizations. In these experiments, two rotational transitions are sequentially pumped on resonance so that the molecular emission or free-induction decay (FID) from a third rotational transition carries the chiral-sensitive molecular signature. This signature exhibits a $\pi$ radian shift between enantiomers in the time domain. Since the FID appears at a frequency that has not been directly excited, this technique provides a way to determine and quantify chirality in a zero background environment. This approach has been applied to a number of chiral molecules such as terpenoids\cite{shubert_identifying_2014, alvin_shubert_enantiomer-sensitive_2014} and alcohols,\cite{patterson_enantiomer-specific_2013,patterson_sensitive_2013, lobsiger_molecular_2015} and it has recently been extended to achieve state-specific population enrichment.\cite{eibenberger_enantiomer-specific_2016,perez_coherent_2017,pratt_pate} All of these molecular systems have a stereogenic center that makes them permanently chiral at the time scales of most experiments.

Another kind of chirality is that arising from the spatial arrangement of a conformer,\cite{zehnacker_suhm} similar to that found in helicenes.\cite{helicenes} In general, any molecule that presents conformational isomerism is posed to become chiral as an equivalent mirror image can be obtained by simple rotations about molecular bonds. In some cases, particularly when the racemization barrier is low, the two forms can constantly and periodically interconvert between one another via tunneling motions through the barrier. Depending on the barrier height, the energy levels may split into two vibrational components (symmetric and antisymmetric), which usually leads to the observation of doublets in the rotational spectrum.\cite{lesarri_propofol} These kinds of internal dynamics provide valuable information on the tunneling rates and potential energy surface of the system under study. They also show that the enantiomeric pairs cannot be isolated. When the interconversion barrier between them is sufficiently high, the two forms equally occupy their corresponding localized potential wells, giving rise to a racemic mixture. This balance can be tilted when the molecule is placed in a chiral environment, such as that created by a nearby permanently\cite{lock-and-key_2007} or transiently\cite{seifert_chiral_2015} chiral molecule. Upon complexation, the chiral molecules preferentially interact with one of the chiral conformers, thereby imprinting a particular chirality.

In this report we experimentally demonstrate state-specific enantiomeric enrichment of chiral conformers based on a M3WM experiment. This builds on the previously demonstrated technique in which a closed cycle of three rotational transitions is simultaneously excited in a phase- and polarization-controlled manner. Here, we generate an ee of our desired enantiomer in a rotational level of choice from the initially racemic ensemble of conformers. Such an ee is then probed by a second M3WM cycle that only produces signal if and only if one of the enantiomers is in excess in a particular rotational state.

Figure 1 shows the racemization barrier profile for the enantiomeric pair of a conformer of cyclohexylmethanol, C$_{7}$H$_{14}$O (CHM). As shown, the racemization barrier is ca. 15 kJ/mol and goes through a transition state structure with C$_{s}$ symmetry. This leads to the formation of both enantiomers in equal amounts in a supersonic expansion. Once they have cooled, they can no longer undergo interconversion or collisional relaxation to lower energy conformations. As with any pair of enantiomers, these two forms have the same moments of inertia, i.e. the same rotational constants A, B and C, and share the same magnitude of the dipole moment components ${\mu}_{a}$, ${\mu}_{b}$ and ${\mu}_{c}$ in the principal axis system. This fact precludes them from being distinguished from one another through traditional rotational spectroscopy as they have identical rotational spectra. However, as illustrated in Figure 1, the two enantiomers have an opposite dipole moment component sign along the b-axis of molecular rotation (yellow in the figure) while the signs of the a-axis dipole moment component (red) and the c-axis dipole moment component (blue) remain unchanged between enantiomers. This difference allows for enantiomer-sensitive rotational spectroscopy to be performed, which in turn permits their differentiation, as well as state-specific population enrichment experiments.

\begin{figure}[t!] 
	\centering
	\includegraphics[width=1\columnwidth]{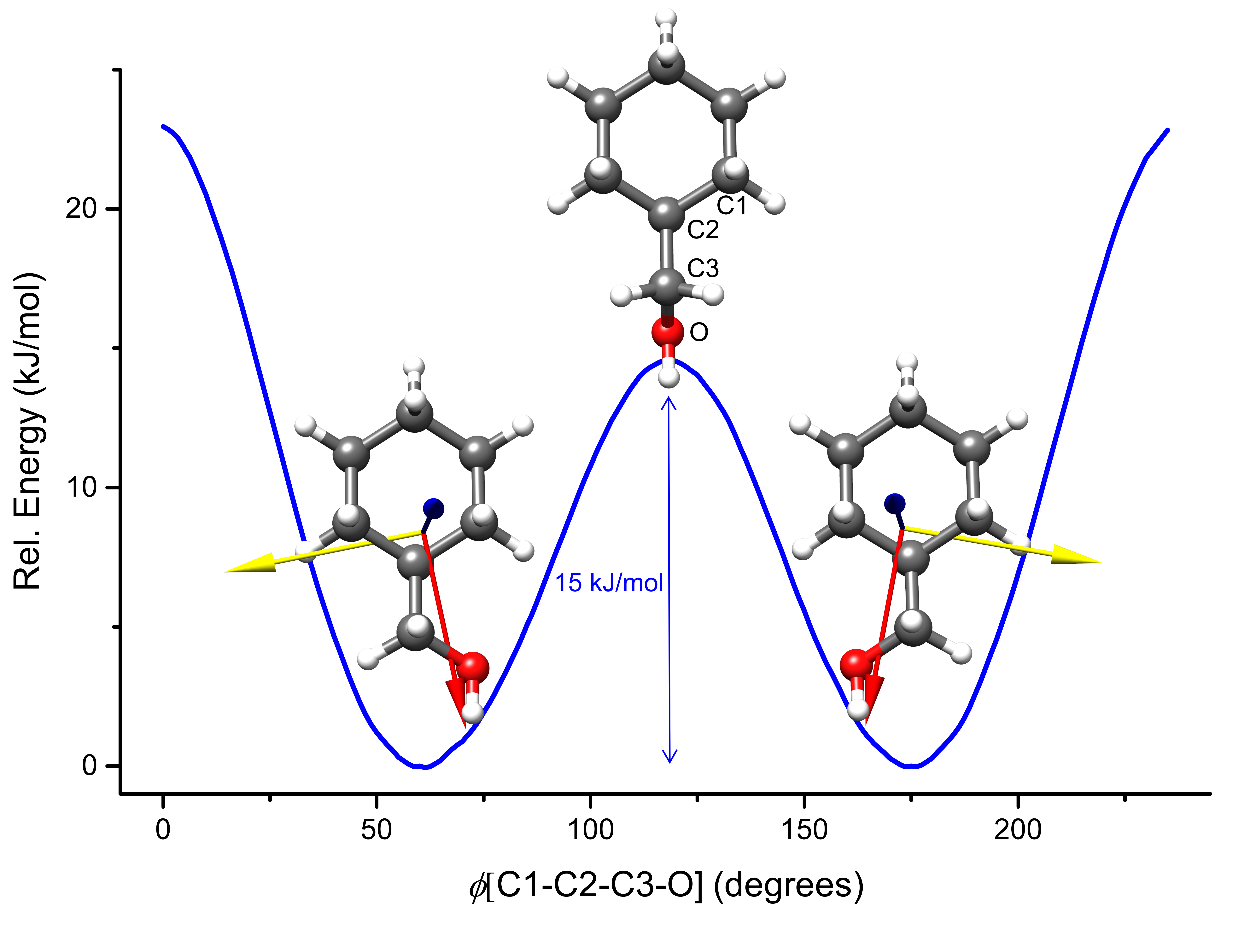} 
	\caption{Full rotational potential energy scan for the interconversion between the two enantiomers of cyclohexylmethanol (CHM) at the MP2/6-311++g(d,p) level of theory. The enantiomeric pair is connected through a transition state structure with C$_{s}$ symmetry that lies 15 kJ/mol higher in energy. The dihedral angle [C1-C2-C3-O] corresponds to the racemization coordinate. The dipole moment components in the principal axis system for the two enantiomers are represented by the colored arrows (a-red, b-yellow, and c-blue, respectively) to show its mirror character. This translates into opposite signs of the triple product $\vec{\mu}_a\cdot(\vec{\mu}_b\times\vec{\mu}_c)$, which is exploited for state-specific population transfer.}
	\label{fig1}
	
\end{figure} 

The experimental setup consists of a chirped-pulse Fourier transform microwave (CP-FTMW)\cite{shubert_identifying_2014} spectrometer with some modifications to allow for chiral-sensitive measurements via M3WM and/or population enrichment.\cite{perez_coherent_2017} A schematic of our setup, as well as more details of the optimal experimental implementation can be found in the supplementary materials, but a concise description of the instrument is presented here. A two-channel 70000 series Tektronix arbitrary waveform generator was used to create a pulse sequence that included first the population transfer pulse sequence (drive, twist, and transfer pulses), followed by a traditional drive and twist pulse sequence for the M3WM portion of the experiment. This pulse sequence was then broadcast across the chamber using a dual polarization Q-Par horn antenna for the drive and population transfer pulses and a horn antenna operating from 1 to 10 GHz for the twist pulses. This second horn was used instead of radio frequency plates to increase the versatility of our instrument and widen the range of target molecules. These pulses interacted with a supersonically expanded sample, and afterwards the free induction decay (FID) of the sample was collected on a fast oscilloscope, averaged, and Fourier transformed. 
Figure 1 shows the molecule CHM chosen to validate our method. It was selected for the ease with which it vaporizes and the fact that it has sufficient transition intensity within the frequency range of our instrument. The sample was heated to 70 $^\circ$C, and the broadband rotational spectrum was measured in the 2-8 GHz frequency region. From the analysis of the obtained rotational spectrum, the rotational constants were determined and are reported in the supplementary materials. Subsequently, several closed loops of transitions were tested to find those that best accommodated the experimental requirements regarding frequency range and polarization. Figure 2 shows the energy levels of CHM involved in the experiment reported here. The rotational levels involved in each transition are denoted using the standard asymmetric top notation, J$ _{K_{a}K_{c}}$, where J is the total rotational angular momentum quantum number and K$_{a}$, K$_{c}$ represent the quantum numbers for the projection of the angular momentum onto the symmetry axis (a- or c-axis) in the two limiting cases of prolate and oblate symmetric tops, respectively.

\begin{figure}[t!] 
	\centering
	\includegraphics[width=1\columnwidth]{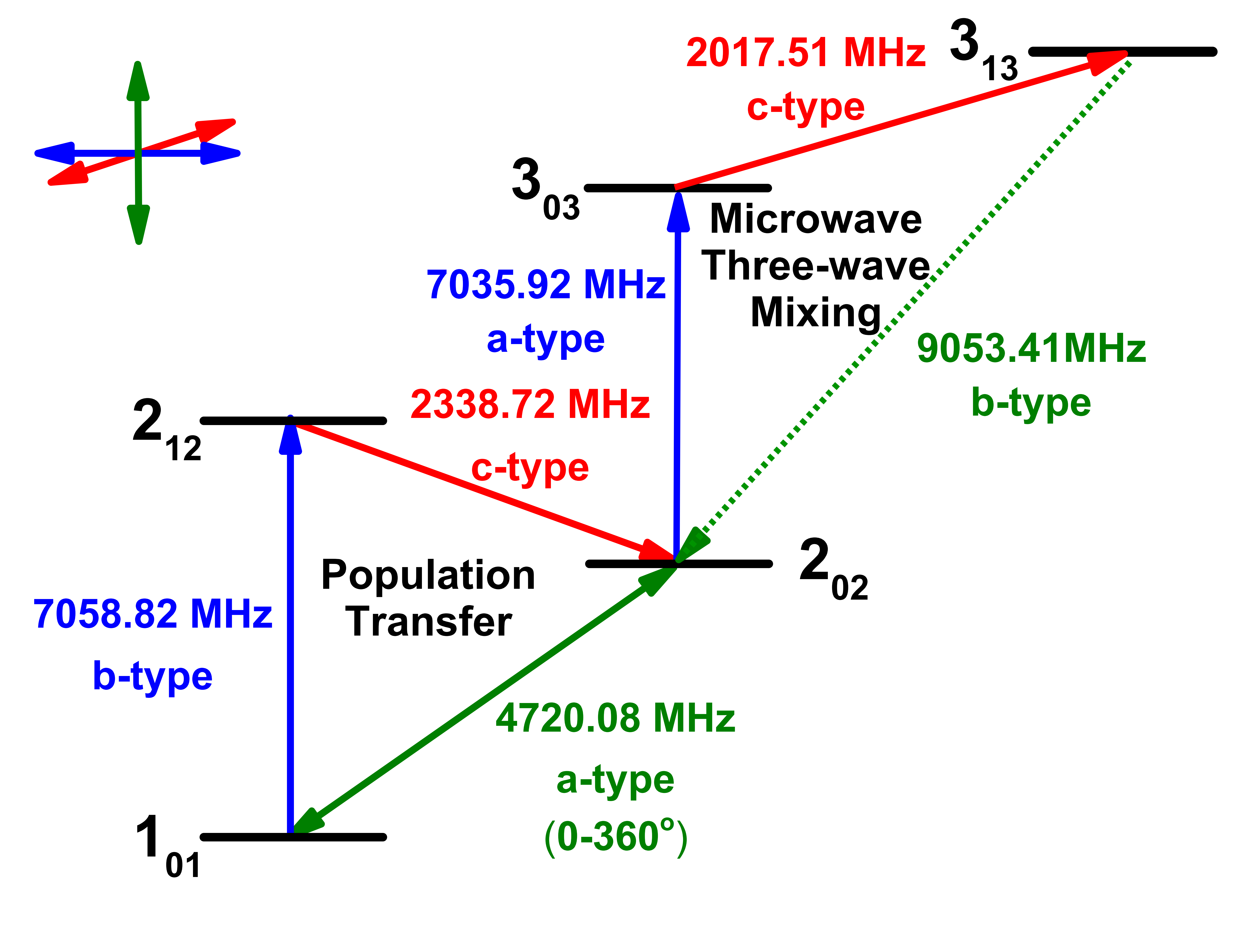} 
	\caption{Relevant rotational energy levels of molecule CHM used for realizing state-specific enantiomer enrichment. The first loop of transitions on the left (population transfer) creates an enantiomer-dependent population difference between the $|2_{02}\rangle$ and $|1_{01}\rangle$ rotational energy levels by varying the relative phase of the transfer pulse at 4720.08 MHz. This induces an enantiomeric excess in both levels, which is then probed by the second three-wave mixing cycle shown on the right (microwave three-wave mixing). The pulses at 7035.92 MHz (“drive”) and 2017.51 MHz (“twist”) create a measurable coherence between the states $|3_{13}\rangle$ and $|2_{02}\rangle$ that is detected in the time domain. See text for more details. Both loops of rotational states consist of all three types of dipole-allowed rotational transitions (a-, b-, and c-type). The color code indicates the orientation of the applied electric fields.  In order to ensure phase stability and reproducibility, all of the microwave pulses where generated from the same source.}
	\label{fig2}	
\end{figure}

In a simple picture, the experimental approach presented here can be seen as two M3WM events performed back-to-back. The first loop of transitions is aimed at creating an ee in both the $|1_{01}\rangle$ and the $|2_{02}\rangle$ rotational levels via cyclic population transfer (CPT) as it has been recently demonstrated. This translates into an excess of one enantiomer in the $|1_{01}\rangle$ state while its mirror image is in excess in the $|2_{02}\rangle$ state or \textit{vice versa}. This ee can then be probed with on a traditional M3WM experiment. The first pulse, the “drive” pulse (blue), at 7058.82 MHz is resonant with the $|2_{12}\rangle \rightarrow |1_{01}\rangle$ transition (driven by ${\mu}_{b}$) and converts the thermal population difference into a coherence. The second pulse, the “twist” pulse (red), at 2338.72 MHz  is resonant with the $|2_{12}\rangle \rightarrow |2_{02}\rangle$ transition (driven by ${\mu}_{c}$) and transfers the initial coherence into a second coherence between the $|2_{02}\rangle$ and $|1_{01}\rangle$ rotational levels (driven by ${\mu}_{a}$). Lastly, a third pulse, the “transfer” pulse (green), at 4720.08 MHz converts the coherence between $|2_{02}\rangle \rightarrow |1_{01}\rangle$ back into a population difference between the $|2_{02}\rangle$ and $|1_{01}\rangle$ rotational states. This population difference is opposite between enantiomers. As one enantiomer is promoted to the upper rotational state, the other is simultaneously transferred to the lower level, creating a state-specific ee from a purely racemic sample. Depending on the relative phases of the applied pulses, the enantiomer of choice can be selectively transferred to the upper or lower rotational levels. This ee is then measured through a traditional M3WM cycle that contains either $|1_{01}\rangle$ or $|2_{02}\rangle$. As shown in Figure 2, we used the latter energy level for our experiment. Similar results can be achieved using the former energy level.
To efficiently drive these processes, careful optimization of the applied pulses must take place beforehand. The corresponding nutation curves (intensity vs pulse duration) were measured. The optimal pulse durations are reported in the supplementary materials, and they were set to maximize the molecular response in the same fashion as previously reported.\cite{perez_coherent_2017} To achieve CPT, the three pulses must be broadcast in three mutually orthogonal directions as indicated by the color code in Figure 2. Afterwards, once an ee has been created, the induced enantiomer-dependent population of the no longer racemic molecular ensemble can be evaluated by applying two additional pulses as in a conventional M3WM experiment. This corresponds to a pulse resonant with the $|3_{03}\rangle \rightarrow |2_{02}\rangle$ rotational transition (driven by  ${\mu}_{a}$) at 7035.92 MHz and the $|3_{13}\rangle \rightarrow |3_{03}\rangle$ transition at 2017.51 MHz (driven by  ${\mu}_{c}$). This second cycle is aimed at creating a measurable enantiomer-specific coherence between the $|3_{13}\rangle$ and $|2_{02}\rangle$ rotational states at 9053.41 MHz (driven by  ${\mu}_{b}$). The FID at this “listen” frequency will show a ${\pi}$ rad phase shift between enantiomers like in a M3WM experiment. As stated above, one of the most important features of M3WM is that it uniquely generates molecular responses for a chiral sample when an ee is present. This is due to the interferometric character of the molecular signal. When the amounts of both enantiomers are identical, a completely destructive interference takes place yielding zero molecular signal. Therefore, the intensity of the listen transition will only be non-zero when an enantiomer-selective population enrichment in the $|2_{02}\rangle$ rotational level is achieved. This can be realized by systematically varying the relative phases between the applied pulses. In our case, we systematically change the phase of the transfer pulse at 4720.08 MHz in 18 degree steps from 0$^\circ$ to 360$^\circ$.
\begin{figure}[t!] 
	\centering
	\includegraphics[width=1\columnwidth]{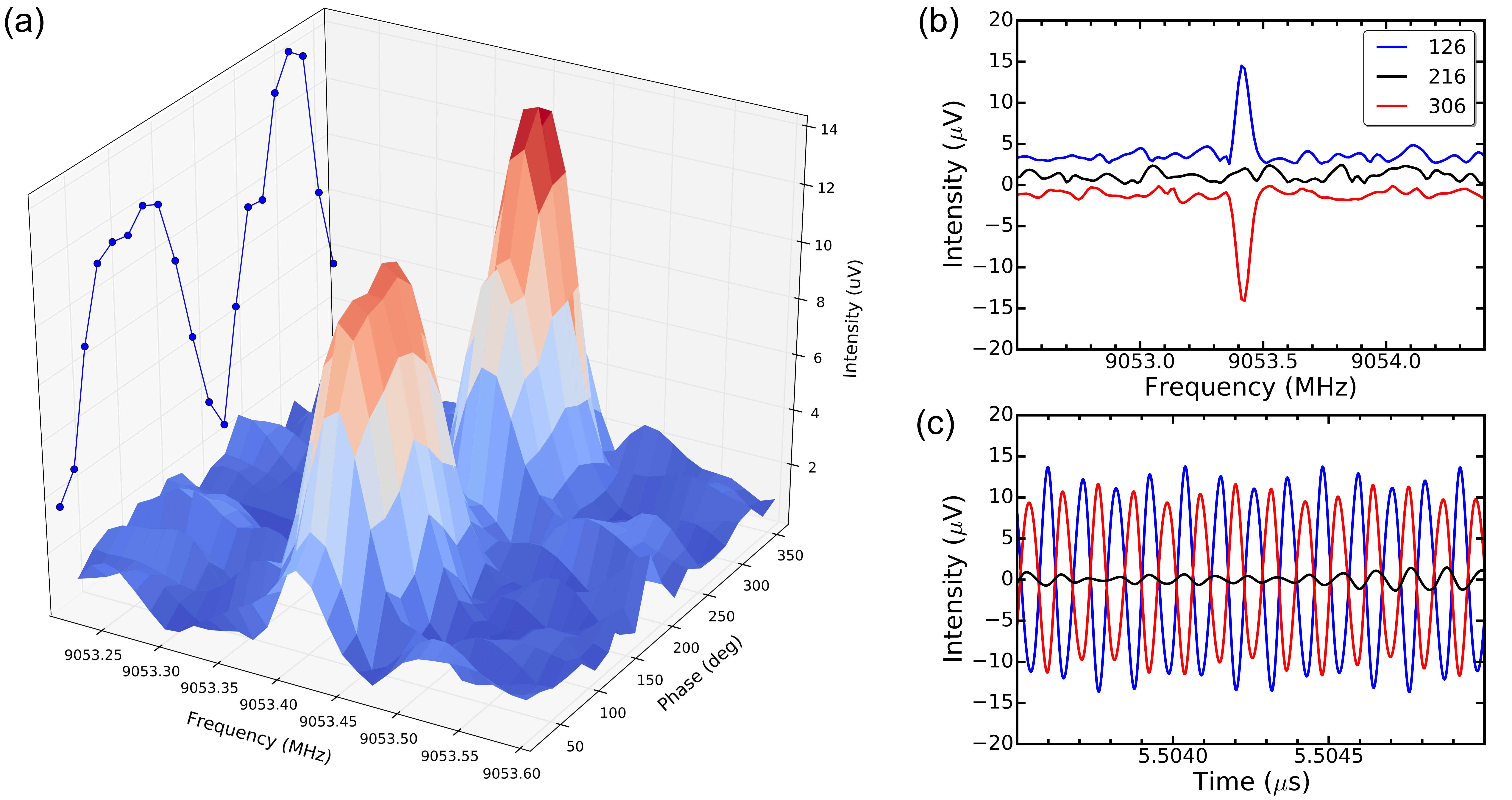} 
	\caption{Experimental demonstration of the state-specific population enrichment of CHM. The molecular response is traced as the magnitude and phase of the $|3_{13}\rangle \rightarrow |2_{02}\rangle$ rotational transition at 9053.41 MHz is changed. Panel (a) shows the signal variation of such a transition when the phase of the $|2_{02}\rangle \rightarrow |1_{01}\rangle$ at 4720.08 MHz is varied in 18 degree steps. Each phase was collected and averaged 100000 times before moving to the next relative phase. Maximum signal was observed at 126 and 306 degree, repectively. This corresponds to the two enantiomers selectively populating or depopulating the $|2_{02}\rangle$ rotational level. Panel (b) shows the frequency domain response at the relevant relative phases (red and blue). The intermediate point at 216 degree (black trace) exhibits no signal as no population enrichment takes place for that particular phase. For the sake of clarity the red trace is shown in the negative scale and a small offset was added. In panel (c) a portion of the corresponding FIDs in the time domain are shown for each phase. As expected, the phases of the observed signals at the two maxima are opposite (${\pi}$ rad phase shift between enantiomers), which further confirms that the enantiomeric pair of CHM is being selectively manipulated.}
	\label{fig3}
\end{figure} 

The results from this five-pulse sequence are shown in Figure 3. The molecular response at 9053.41 MHz was recorded in the time domain for 40 ${\mu}$s as a function of the transfer pulse phase. As shown, we observe that the amplitude of this transition exhibits two maxima separated by 180 degree at 126 and 306 degree, respectively. This corresponds to the two relative phases at which the enantiomer-specific enrichment in $|2_{02}\rangle$ is optimal. Note that this provides a very efficient, robust way for selecting the enantiomer to be enriched by simply tuning the transfer pulse to one of those relative phases. When the population transfer does not take place, in this case at 216$^\circ$, there is no induced ee in the $|2_{02}\rangle$ level and the second three-wave mixing cycle yields no signal as shown in panel (b) of Figure 3. What is more, in order to further corroborate that the enantiomers are actually being selectively transferred to the $|2_{02}\rangle$ rotational level, panel (c) shows a portion of the enantiomer-dependent FID at the three relevant relative phases mentioned above. As expected, the opposite enantiomers are 180 degree out-of-phase (red and blue traces), while the intermediate phase at 216 degree shows no molecular response. This confirms that, in fact, the two enantiomers of CHM are being selectively transferred to the target $|2_{02}\rangle$ rotational level.

The current results represent the first experimental demonstration of enantiomer-specific population enrichment in a pair of chiral conformers. We have shown that the initially racemic mixture of enantiomers can be manipulated to yield an ee of one of the enantiomers by reducing or increasing the population of its counterpart as well as doing the opposite to the population of the desired enantiomer in a particular rotational level. The identity of the desired enantiomer can be easily swapped by simply changing the relative phase of one of the applied pulses. As the obtained ee is dependent upon the initial populations between rotational levels, we envision that a clear enhancement will be realized using more sophisticated, tailored pulse sequences and/or evolutionary algorithms are applied. Our approach takes advantage of all of the well-known features of rotational spectroscopy, and it is, therefore, particularly well-suited for the analysis of complex mixtures. As this has been realized, further experiments targeted at a spatial separation can be performed. One proposed experiment to investigate would be using electrostatic deflectors for separating the two enantiomers based on their dipole-moment-to-mass ratios and the different effective dipole moment in a particular rotational state.\cite{filsinger_pure_2009,filsinger_selector_2008}. The use of inhomogeneous electric fields has been applied to deflect, focus, decelerate or/and accelerate molecules in a molecular beam.\cite{meerakker_taming_2008,trippel_kupper} The successful combination of such kinds of experiments with the enantioenrichment of rotational levels currently presented promises to be able to generate enantiopure molecular beams of choice that would enable a new class of unprecedented ultra-high resolution experiments as well as enantiomer- and state-selective collision experiments. Alternatively, the sample can be further enriched by applying laser pulses to promote the desired enantiomer to higher, less populated vibrational or electronic levels to achieve a complete depletion. Moreover, our results open the door to study other systems of particular interest such as molecules with a relatively low racemization barrier, especially for those in which parity violation dominates over tunneling interconversion in the ground state as HSSSH.\cite{fabri_tunneling_2015,berger_parity_2001} With further developments, our results have the potential to provide chemists with additional tools for discriminating chirality and gaining more insight on parity violation, and what is more, ultimately choose the handedness of a chemical system.







\begin{acknowledgement}
This work was supported by the Sonderforschungsbereich 1319 “Extreme light for sensing and driving molecular chirality (ELCH). C.P. acknowledges a grant by the Alexander von Humboldt Stiftung. A.L.S. acknowledges the Louise Johnson Fellowship of the excellence cluster “The Hamburg Centre for Ultrafast Imaging", supported by the Deutsche Forschungsgemeinschaft. All the authors thank Sergio R. Domingos for fruitful discussions. 
\end{acknowledgement}



\begin{suppinfo}
Schematic of our experiment. Optimal experimental conditions. Rotational constants and list of observed frequencies of cyclohexylmethanol. 
\end{suppinfo}

\bibliographystyle{achemso}
\bibliography{references}

\end{document}